\documentclass[12pt]{article}

\begin{document}

\title{On the entropy of spanning trees \\
on a large triangular lattice}
\author{ M. L. Glasser\\
Department of Physics, Clarkson University \\
Potsdam, New York 13699\\
\\
and\\
\\
F. Y. Wu \\
 Department of Physics,
Northeastern University\\
 Boston, Massachusetts 02115}

%\pacs{05.50.+q}

\maketitle

\begin{abstract}
The double integral representing the entropy $S_{\rm tri}$ of
spanning trees on a large triangular lattice
 is evaluated using two different
methods, one algebraic and one graphical.  Both methods lead to
the same result

\begin{eqnarray}
S_{\rm tri} &=& (4\pi^2)^{-1} \int_0^{2\pi}d\theta
\int_0^{2\pi}d\phi
  \ln \big[6-2\cos \theta - 2\cos \phi -2 \cos({\theta+\phi} ) \big] \nonumber \\
 &=& \big(3\sqrt 3/\pi\big)  \big( 1 -  5^{-2} +  7^{-2}   - 11^{-2} +13^{-2}
  - \cdots \big). \nonumber
\end{eqnarray}
\end{abstract}

\newpage
\section{Introduction}
It is well-known that spanning trees on an $n$-vertex graph $G$
can be enumerated by computing the eigenvalues of the Laplacian
matrix ${\bf Q}$ associated with $G$ as
\begin{equation}
N_{\rm ST}(G) = \frac 1 n \prod_{i=1}^{n-1} \lambda_i\ ,
\label{spanningtree}
\end{equation}
where $\lambda_i,\ i=1,2,\cdots,n-1$, are the $n-1$ nonzero
eigenvalues of ${\bf Q}$
 \cite{biggs}.
In physics one often deals with lattices.
 For large lattices the number of spanning trees $N_{ST}$ grows
exponentially in $n$.   This permits one to  define the entropy of
spanning trees for a class of lattices $G$ as the limiting value
\begin{equation}
S_G = \lim_{n\to \infty} \frac 1 n \ln N_{\rm ST}(G).
\label{entropy}
\end{equation}
The resulting finite number  $S_G$ is a quantity of physical
interest.

\medskip
For regular lattices in $d$ dimensions
 the eigenvalues $\lambda_i$  can be computed by standard
means \cite{tzengwu} leading to expressions of $S_G$ in terms of
$d$-dimensional definite integrals. The resulting integrals, which
are independent of boundary conditions, have been given  for a
number of regular lattices  \cite{shrockwu}, but very few of the
definite integrals have been evaluated in closed forms.

\medskip
To be sure, the entropy for the square lattice has been known and
computed
 \cite{temp}. It is found to be
\begin{eqnarray}
S_{\rm sq} &=& (4\pi^2)^{-1} \int_0^{2\pi}d\theta \int_0^{2\pi}d\phi
  \ln (4-2\cos \theta - 2\cos \phi) \nonumber \\
  &=& \frac {4} \pi \;{\rm Gl}_2 \bigg( \frac \pi 2 \bigg) \nonumber \\
&=& \frac {4} \pi \bigg[ 1 - \frac 1 {3^2} + \frac 1 {5^2}
   -\frac 1 {7^2} +\frac 1 {9^2} -\cdots \bigg] \nonumber \\
&=&\frac{4}{\pi}\;{\bf{G}}\nonumber \\
  &=&  1.166\ 243\ 616\ ...  \label{sq}
\end{eqnarray}
where
\begin{equation}
{\rm Cl}_2(\theta)=\sum_{n=1}^\infty \frac  {\sin(n\theta)}  {n^2}
\label{Clausen}
\end{equation}
 is the Clausen's function and
  ${\bf{G}}$ is the Catalan constant. The algebra reducing the
double integral in (\ref{sq}) into the series given in the third
line is straightforward and can be found in \cite{kas,fisher}.

\medskip
The only other published closed-form numerical result on the
entropy of spanning
 trees is that of  the triangular lattice
\cite{wu} for which one has
 \begin{eqnarray}
S_{\rm tri} &=& (4\pi^2)^{-1} \int_0^{2\pi}d\theta
\int_0^{2\pi}d\phi
  \ln \big[6-2\cos \theta - 2\cos \phi -2 \cos({\theta+\phi} ) \big] \nonumber \\
 &=& \frac{5}{\pi}\ {\rm Cl}_2\bigg(\frac \pi 3\bigg) \nonumber\\
  &=& \frac {3\sqrt 3} \pi \bigg[ 1 - \frac 1 {5^2} + \frac 1 {7^2}
 -\frac 1 {11^2} +\frac 1 {13^2}
  - \cdots \bigg] \nonumber \\
    &=& 1.615\ 329\ 736\ 097... \ . \label{tri}
\end{eqnarray}
 However,
no details leading from the double integral in (\ref{tri}) to the
final series expression have been published. The expression in the second
line is new.

\medskip
The purpose of this note is to provide some details of the 
steps  leading to the second and third lines in (\ref{tri})
which, as we shall see, are  nontrivial.  
We evaluate the
integral in (\ref{tri}) using two different methods: a direct
evaluation and a graphical approach. We first evaluate the
integral algebraically, and then in an alternate approach we use
graphical considerations to convert the spanning tree problem to a
Potts model which is in turn related to an $F$ model solved by
Baxter \cite{baxter}. In either case, the same  series expression
in (\ref{tri}) is deduced.\footnote{The numerical values of
$S_{\rm tri}$  reported in \cite{wu} and \cite{baxter} contain typos
in the last two digits.}

\section{Algebraic approach}
In this section we evaluate the integral (\ref{tri}) directly.

\medskip
Using the integration formula
\begin{eqnarray}
\frac 1 {2\pi} \int_0^{2\pi} d\phi
  \ln \Big[ 2A + 2B \cos \phi + 2C \sin \phi \Big] = \ln \Big[ A +
   \sqrt {A^2-B^2-C^2} \Big], \nonumber
\end{eqnarray}
one of the two integrations in (\ref{tri}) can be carried out,  giving
 \begin{equation}
S_{\rm tri} = \frac 1 \pi\int_0^\pi d\theta\
\ln\bigg[3-\cos\theta+\sqrt{(7-\cos\theta)(1-\cos\theta)}\bigg].\label{triI}
\end{equation}
 Next, we use the identity
\begin{equation}
3-\cos\theta+\sqrt{(7-\cos\theta)(1-\cos\theta)}= \frac 2 3
\bigg(v+\sqrt{3+v^2}\bigg)\bigg( 2v +\sqrt{3+v^2}\bigg)
\end{equation}
where $v=\sin (\theta/2)$.  Then  (\ref{triI}) becomes
\begin{equation}
S_{\rm tri} = \ln \bigg(\frac 2 3 \bigg)+I_1 +I_2,\label{II}
\end{equation}
where
\begin{equation}
I_m= \frac 2 \pi \int_0^{\pi/2} d\theta\ \ln \bigg(m\sin \theta
       + \sqrt {3+\sin^2 \theta} \bigg) , \quad m=1,2 \label{II1}
\end{equation}
after  a change of variable $\theta/2 \to \theta$.

\medskip
 To evaluate $I_1$, we consider the more general integral
\begin{eqnarray}
f(a) =\frac 2 \pi \int_0^{\pi/2} \ln \bigg(\sin \theta
      + \sqrt {a^2+\sin^2 \theta} \bigg) \nonumber
\end{eqnarray}
such that $f(\sqrt 3) = I_1$. Using
\begin{equation}
 \frac 2 \pi \int_0^\pi d\theta\ln \sin \theta=\frac 2 \pi \int_0^\pi
 d\theta\ln \cos \theta = -\ln 2,  \label{log}
\end{equation}
 we have
\begin{eqnarray}
f(0) &=&  \frac 2 \pi \int_0^\pi
d\theta\  \ln (2\sin \theta) d\theta=0,\nonumber \\
 f'(a)&=&\frac {2a} {\pi }  \int_0^{\pi/2}
\frac{d\theta}{(\sin \theta +\sqrt{a^2+\sin^2\theta})\sqrt{a^2
  +\sin^2\theta}} \nonumber\\
&=& \frac 2 {\pi a}\tan^{-1} a\ .\nonumber
\end{eqnarray}
The reduction of $f'(a)$ is effected by noting that
$(\sqrt{\sin^2\theta+a^2}+\sin\theta)^{-1}=(\sqrt{\sin^2\theta+a^2}-\sin\theta)/a^2$.
Hence, $ f(a) =( 2/ \pi) {\rm Ti}_2(a)$ and
\begin{eqnarray}
I_1 = \frac 2 \pi\; {\rm Ti}_2 (\sqrt 3) \label{I1}
\end{eqnarray}
where
\begin{eqnarray}
{\rm Ti}_2(a) &=& \int_0^a\frac{\tan^{-1}t}{t}\;dt \label{arctan}\\
&=&  a - \frac {a^3} {3^2} +\frac {a^5} {5^2} - \frac {a^7} {7^2} +\cdots  \nonumber 
% &=&\sum_{n=0}^{\infty}(-1)^n\frac{a^{2n+1}}{(2n+1)^2} \nonumber
\end{eqnarray}
is the inverse tangent integral function described in \cite{lewin}.

\medskip
 To evaluate $I_2$, we introduce the
following  identity
\begin{equation}
2 \sin \theta +\sqrt{3+\sin^2\theta} = \sqrt 3 \cos \theta \sqrt{
  \frac { 1+u(\theta)} { 1-u(\theta)}}, \quad 0\leq \theta \leq \pi/2,
\label{id1}
\end{equation}
where
\begin{eqnarray}
u (\theta)= \frac {2 \sin \theta} { \sqrt{3+\sin^2\theta}}
.\nonumber
\end{eqnarray}
Substituting (\ref{id1}) into $I_2$ in (\ref{II1}) and making use
of (\ref{log}), we obtain
\begin{eqnarray}
I_2 &=& {1\over 2}\ln 3 - \ln 2 +\frac{{1}}{\pi}\int_0^{\pi/2}\ln
\bigg[
  \frac {1+u(\theta)}{1-u(\theta)}\bigg] \ d\theta \nonumber \\
 &=& {1\over 2}\ln 3 - \ln 2
+\frac{\sqrt{3}}{2\pi}\int_0^1 \frac 1 {(1-u^2/4)\sqrt{1-u^2}} \ln
\bigg( \frac {1+u}{1-u} \bigg)\ du
\end{eqnarray}
after an obvious change of integration variable in the last step.

\medskip
Next we express the factor in the integrand as
\begin{eqnarray}
\frac 1 {1-u^2/4}
 = \frac 1 2\bigg[ \frac 1 {1-u/2 } +\frac 1 {1+u/2}\bigg]\nonumber
\end{eqnarray}
and set
\begin{eqnarray}
x= \sqrt {(1-u)/(1+u)}\ .\nonumber
\end{eqnarray}
After a little reduction we obtain
\begin{eqnarray}
I_2 ={1\over 2}\ln 3 - \ln 2
 -\frac {2\sqrt 3} \pi
 \int_0^1\frac{\ln\; x}{3x^2+1}dx -\frac {2\sqrt 3} \pi
\int_0^1\frac{\ln\; x}{x^2+3}dx.\nonumber
\end{eqnarray}
From integration by parts one has
\begin{eqnarray}
\int_0^1\frac{\ln\; x}{x^2+a^2}dx &=& -\bigg( {\frac 1 a}\bigg)
\int_0^1 \frac {dx} x \tan^{-1}
 \bigg(\frac x a \bigg) \nonumber \\
&=& -\bigg(\frac{1}{a}\bigg)\;{\rm Ti}_2\bigg(\frac 1 a \bigg).
\end{eqnarray}
This yields
\begin{eqnarray}
I_2={1\over 2}\ln 3 - \ln 2 +\frac 2 \pi \bigg[{\rm
Ti}_2\big(\sqrt{3}\big)+{\rm Ti}_2\bigg({1\over {\sqrt{3}}}
\bigg)\bigg].  \label{I2}
\end{eqnarray}

To express $I_1$ and $I_2$ in terms of  the Clausen's
series (\ref{Clausen}), we have from Eqs. (2.6), (4.31) and (4.18) of \cite{lewin}
the identities
\begin{eqnarray}
 {\rm Ti}_2\big(y^{-1}\big) &=& {\rm Ti}_2(y) -\frac{\pi}{2}\ln
y, \quad y>0 \label{identity1} \\
{\rm Ti}_2(\tan\theta)&=&\theta\ln(\tan\theta)+\frac{1}{2}
  \bigg[{\rm Cl}_2(2\theta)+{\rm Cl}_2(\pi-2\theta)\bigg],
  \quad \theta>0 \label{identity2}\\
 {\rm Cl}_2\bigg(\frac{2\pi}{3}\bigg)&=&\frac{2}{3}\;{\rm Cl}_2\bigg(\frac{\pi}{3}\bigg).
\label{identity3}
\end{eqnarray}
Setting $y=\sqrt 3$ in (\ref{identity1}) and $\theta = \pi/3$ in (\ref{identity2}), we
obtain  after making use of (\ref{identity3}) 
\begin{eqnarray}
{\rm Ti}_2(\sqrt 3)    &=&  {\rm Cl}_2 \bigg( \frac \pi 3 \bigg) + \frac \pi {6} 
   \ln 3\nonumber \\
{\rm Ti}_2\bigg({1\over\sqrt 3}\bigg)    &=&  {\rm Cl}_2 \bigg( \frac \pi 3 \bigg)
 -\frac \pi {12} \ln 3\ . \label{id}
\end{eqnarray}
Thus,  combining (\ref{I1}) and (\ref{I2}) with (\ref{II}), we obtain  
\begin{equation}
S_{\rm tri} =\frac 5 \pi \;{\rm Cl}_2\bigg( \frac \pi 3 \bigg) \label{trif}
\end{equation}
as given in the second line in (\ref{tri}).

\medskip
To express $S_{\rm tri}$ in the form of the series given in (\ref{tri}), we note that
\begin{equation}
{\rm Cl}_2\bigg( \frac \pi 3 \bigg) =  \frac {\sqrt 3} 2  \;{S} \label{cl}
\end{equation}
where
\begin{eqnarray}
{ S} &=& \sum_{m=0}^\infty \Bigg[ \frac 1 {(6m+1)^2} +\frac 1 {(6m+2)^2} 
         -\frac 1 {(6m+4)^2} -\frac 1 {(6m+5)^2} \Bigg] \nonumber \\
 &=& S_1 + \frac 1 {2^2}\; S_2 \nonumber 
\end{eqnarray}
with
\begin{eqnarray}
 { S}_1 &=& \sum_{m=0}^\infty \Bigg[ \frac 1 {(6m+1)^2} 
          -\frac 1 {(6m+5)^2} \Bigg] ,\nonumber \\
{ S}_2 &=& \sum_{m=0}^\infty \Bigg[ \frac 1 {(3m+1)^2} - \frac 1 {(3m+2)^2} \Bigg]. \nonumber
\end{eqnarray}
Separating terms in $S_2$ with odd and even denominators,   we have
\begin{eqnarray}
S_2 = S_1 - \frac 1 {2^2}\; S_2 \; . \nonumber
\end{eqnarray}
It follows that  $S_2 =  (4/ 5) S_1$ hence $S=(6/5)S_1$, 
and from (\ref{trif}) and (\ref{cl}) we obtain
\begin{eqnarray}
S_{\rm tri} &=&\frac {3\sqrt 3} \pi \; S_1  \nonumber \\
  &=& \frac {3\sqrt 3} \pi \; \bigg[ 1 - \frac 1 {5^2} + \frac 1 {7^2}
 -\frac 1 {11^2} +\frac 1 {13^2}
  - \cdots \bigg]. \nonumber 
\end{eqnarray}
This is the result given in (\ref{tri}).

\section{Graphical approach}
In this section we evaluate the integral (\ref{tri}) with the help
of a graphical mapping.

\medskip
   The medial graph (or the surrounding lattice) ${\cal L}$' of
a triangular lattice ${\cal L}$ is the kagome lattice shown in
Fig. 1, where we have shaded its triangular faces  for
convenience.
 We start from a well-known
equivalence between the $q$-state Potts model on ${\cal  L}$ and
an ice-rule vertex model on ${\cal L}$' \cite{tl} -
\cite{wupotts}. Denote the respective partition functions by $Z_{\rm
Potts}$ and $Z_{\rm kag-ice}$.  Then
 the equivalence reads
\begin{equation}
Z_{\rm Potts}(q, \sqrt q \ x) = q^{n/2} Z_{\rm kag-ice}(1,1,x,x,
   t^{-1} + x t^2, t+xt^{-2})
\label{e1}
\end{equation}
where
\begin{eqnarray}
x &=& (e^K-1)/\sqrt q\ , \nonumber \\
t^3 + t^{-3} &=& \sqrt q\ ,
\end{eqnarray}
 $K$ being the nearest-neighbor interaction of the Potts model,
 and the arguments of $Z_{\rm kag-ice}$ are
weights\footnote{In comparing vertex  weights with those in \cite{wupotts},
it should be noted that for the triangular lattice which we are
considering,
 roles of  shaded and unshaded faces
 are reversed from those indicated in \cite{wupotts}.}
    of the ice-rule model   as depicted in Fig. 2.

\medskip
On the other hand, it is also known that by taking an appropriate
$q\to 0$ limit the Potts partition function generates spanning
trees \cite{wu,fk}. Write the spanning tree generating function in
a slightly more general form by attaching a weight $x$ to each
line segment of the tree.  Since the number of lines in a spanning
tree on an $n$-vertex graph is fixed at $n-1$, this merely adds an
overall factor $x^{n-1}$. Then, the  equivalence reads
\cite{wu,fk}
\begin{equation}
x^{n-1} N_{\rm ST} = \lim_{q\to 0} q^{-(n+1)/2}Z_{\rm Potts}(q,
\sqrt q \ x).\label{e2}
\end{equation}
The equivalences (\ref{e1}) and (\ref{e2}) hold quite generally
for finite lattices, provided that appropriate vertex weights are
defined for the boundary \cite{bkw}.

\medskip
We next combine  (\ref{e1}) and (\ref{e2}) and take the
thermodynamic limit $n\to \infty$.
 Assuming the
two limits on $n$ and $q$ commute, this leads to
\begin{equation}
S_{\rm tri} = - \ln x + z_{\rm kag-ice} (1,1,x,x,t^{-1} + x t^2,
t+xt^{-2}), \quad t=e^{i\pi/6} \label{e3}
\end{equation}
where
\begin{eqnarray}
z_{\rm kag-ice}=\lim_{n\to\infty} n^{-1} \ln Z_{\rm kag-ice}.
\nonumber
\end{eqnarray}
 We note in passing that the vertex weights in (\ref{e3})
now  satisfy the free-fermion condition \cite{fanwu}
\begin{eqnarray}
1\cdot 1 + x\cdot x = (t^{-1} + x t^2)\cdot ( t+xt^{-2}) \nonumber
\end{eqnarray}
so $z_{\rm kag-ice}$ can be evaluated by standard means such as
using Pfaffians
 \cite{wu}.
This verifies  the integral expression of $S_{\rm tri}$ given in
(\ref{tri}).

\medskip
Alternately, using a mapping introduced by Lin \cite{lin} (see
also \cite{baxter1}), the kagome ice-rule model can be mapped into
an ice-rule model on the triangular lattice. The trick is to
'shrink' each shaded  face of the kagome lattice into a point as
shown in Fig. 1. This results in  a  20-vertex ice-rule model on
the triangular lattice which has 3 arrows in and 3 arrows out at
every vertex.

\medskip
There are 4 different vertex types in the 20-vertex model:
  (i) six vertices
in which the 3 incoming arrows are adjacent, (ii) two vertices in
which the incoming and outgoing arrows alternate around the
vertex, (iii) six vertices in which 2 incoming arrows are adjacent
while the third in-arrow points in a direction opposite to {\it one} of
the 2 adjacent incoming arrows, and (iv) six vertices in which 2
incoming arrows are adjacent while the third  in-arrow points
opposite to the {\it other}  adjacent incoming arrow.  These 4
configurations  are shown in the left-side column in Fig. 3 to which
we assign respective energies $\epsilon_i,\ i=1,2,3,4$.  The other 16
ice-rule configurations are obtained by rotations.

\medskip
The shrinking processes from which these ice-rule configurations
 are deduced are also shown in Fig. 3.
Write the 4 weights as $a,b,c,c'$, respectively.  We read off from
Fig. 3 to obtain
\begin{eqnarray}
a&=&e^{-\epsilon_1/kT}  = 1\cdot 1\cdot x \nonumber\\
b&=&e^{-\epsilon_2/kT} = (t+xt^{-2})^3 + x^3  \nonumber \\
c&=&e^{-\epsilon_3/kT} = x(t+xt^{-2})^2 +x^2(t^{-1}+xt^2) \nonumber \\
c'&=&e^{-\epsilon_4/kT} =  x^2(t+xt^{-2}) +x(t^{-1}+xt^2)^2 \ .
\label{32}
\end{eqnarray}
 Thus, we have established the equivalence
\begin{eqnarray}
z_{\rm kag-ice} (1,1,x,x,t^{-1} + x t^2, t+xt^{-2}) = z_{\rm
tri-ice} (a,b,c,c') ,\label{e4}
\end{eqnarray}
where $z_{\rm tri-ice}$ is defined similar to $z_{\rm kag-ice}$.

\medskip
Baxter \cite{baxter} has considered the evaluation of $z_{\rm
tri-ice}$ in the subspace of $c=c'$.  Now the right-hand side of
(\ref{e3}) is independent of the value of $x$ so we have
the freedom to  choose a
value of $x$ for which $z_{\rm ice-tri}$ can be evaluated. It is
readily verified that by taking
 $x=1/\sqrt 3$ and
$t=e^{i\pi /6}$  we have
\begin{eqnarray}
a= 1/\sqrt 3, \quad b=3a, \quad c=c'=2a \nonumber
\end{eqnarray}
 which is a case solved by Baxter.  In this case Baxter  obtained 
 \begin{eqnarray}
 z_{\rm tri-ice}(a, 3a, 2a, 2a) 
&=& \ln a 
+ P \int_{-\infty}^\infty \frac {dx} x\; \frac {1+e^{-x}}
  { (e^{x} -1 + e^{-x})(1+ e^{-2x} + e^{-4x})} \nonumber \\
&=&  \ln a +
\frac {3\sqrt 3} \pi \bigg[ 1 - \frac 1 {5^2} + \frac 1 {7^2}
 -\frac 1 {11^2} +\frac 1 {13^2}
  - \cdots \bigg], \label{e5}
 \end{eqnarray}
where $P$ denotes the principal value.
Combining (\ref{e3}) and (\ref{e4}) with (\ref{e5}) and
$a=1/\sqrt 3$,
 we obtain
\begin{eqnarray}
S_{\rm tri} =\frac {3\sqrt 3} \pi \bigg[ 1 - \frac 1 {5^2} + \frac 1
{7^2}
  -\frac 1 {11^2} +\frac 1 {13^2}  - \cdots \bigg].  \nonumber
\end{eqnarray}
This is the result given in (\ref{tri}) as first reported in
\cite{wu}.

\section*{Acknowledgment}
Work has been supported in part by NSF Grants DMR-0121146 (MLG)
and DMR-9980440 (FYW). We thank W. T. Lu for assistance in
preparing the graphs.

\newpage

\begin{center}
{\bf Figure captions}
\end{center}

\noindent
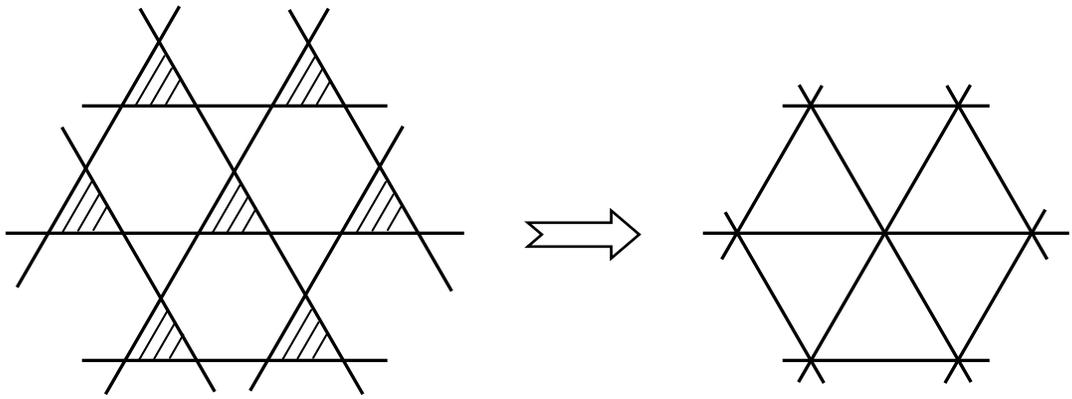
Fig. 1.  The shrinking of a kagome lattice into a triangular lattice.  Each
shaded triangular face is shrunk into a point.
\vskip .5cm

\noindent
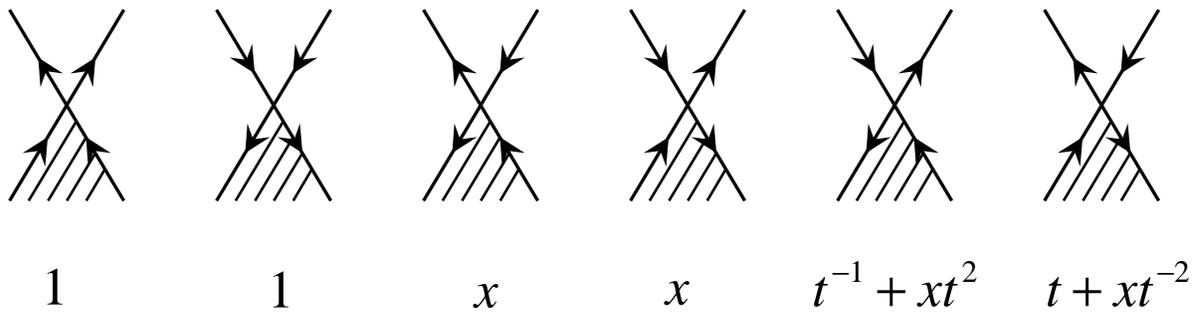
Fig. 2. The six ice-rule configurations of the kagome lattice and the associated
vertex weights.

\vskip .5cm
\noindent
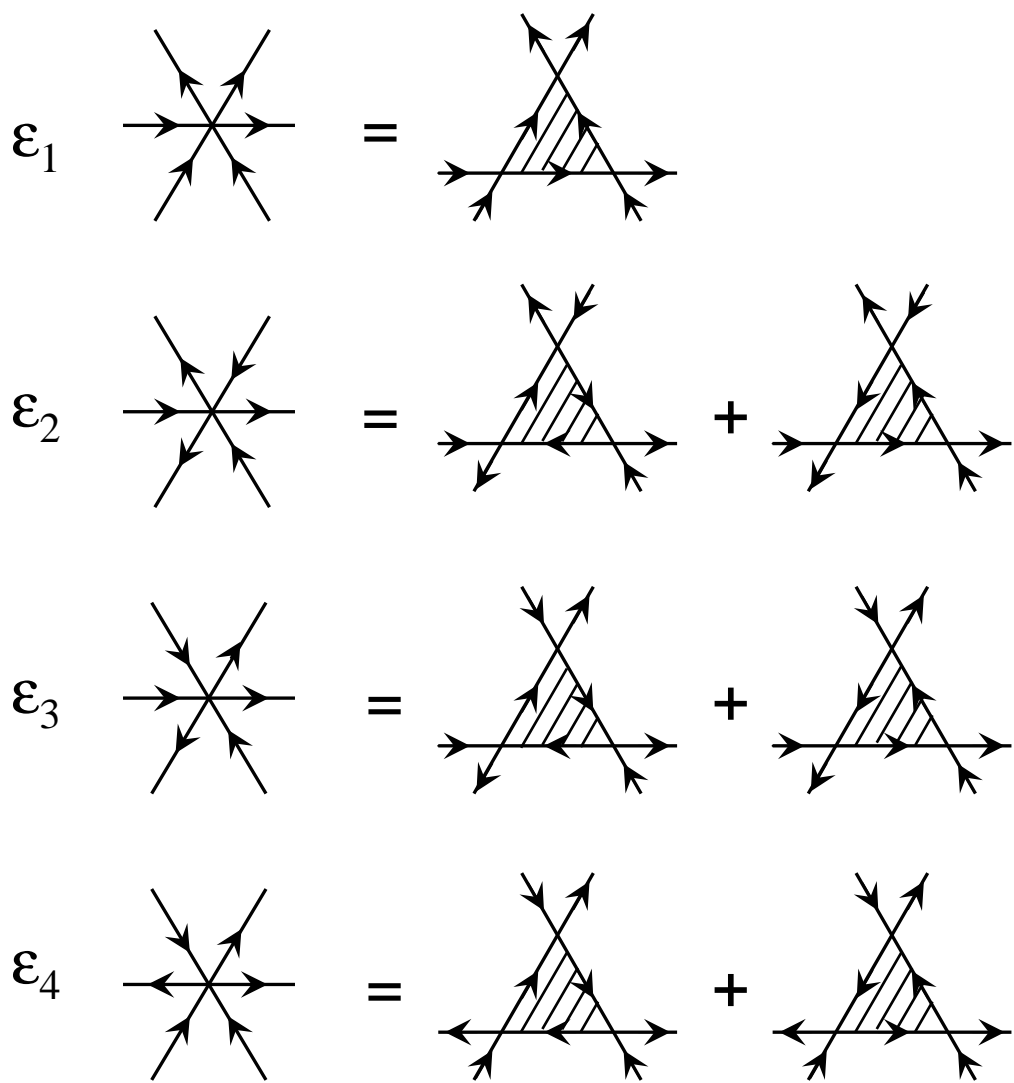
Fig. 3.  The four types of ice-rule configurations of the triangular
lattice and the associated shrinking processes.

\end{document}